\documentclass[12pt]{iopart}
\input epsf

\newcommand{\beq}{\begin{equation}}
\newcommand{\eeq}{\end{equation}}
\newcommand{\beqa}{\begin{eqnarray}}
\newcommand{\eeqa}{\end{eqnarray}}
\newcommand{\non}{\nonumber}

\begin{document}

\title{A new approach for the analytic computation of the Instantaneous 
Normal Modes spectrum }

\author{Andrea Cavagna \dag
\footnote[1]{ Current address: {\it Department of Physics and Astronomy,
The University of Manchester, Oxford Road, Manchester, M13 9PL, UK}.
E-mail: andrea@a13.ph.man.ac.uk.},
Irene Giardina \dag\footnote[2]{
Current address: {\it Service de Physique Theorique 
CEA-Saclay, Orme des Merisiers, 91191 Gif sur Yvette Cedex, France}.
E-mail: giardina@spht.saclay.cea.fr.} and 
Giorgio Parisi \ddag\footnote[3]{
E-mail: giorgio.parisi@roma1.infn.it}
}

\address{\dag\ Theoretical Physics, University of Oxford,
1 Keble Road, Oxford, OX1 3NP, United Kingdom}

\address{\ddag\ Dipartimento di Fisica, Universit\`a di Roma La Sapienza and
INFN Sezione di Roma I, P.le Aldo Moro 5, 00185 Roma, Italy}

\begin{abstract}
{\sl 
In the context of the Instantaneous Normal Mode approach,
the spectrum of the Hessian of Hamiltonian  is a key quantity
to describe liquids behaviour. The determination of the
spectrum represents a major task for theoretical studies, and has been
addressed recently in various works.
In this work a new approach for the analytic computation of the Hessian
spectrum is presented. The one dimensional case for a system
of particles interacting via a purely repulsive potential at low density 
is analyzed in details and the spectrum is computed exactly also in the 
localized sector. Finally, the possible  extensions of the method are 
discussed,
together  with a comparison with different  approaches to  the problem.
}
\end{abstract}


\submitted{{\noindent \it }}

\section{Introduction}

In this work we  describe an approach  we have recently
developed to  compute analytically  the so called
Instantaneous Normal Mode (INM) spectrum of a liquid system
\cite{noispettro}. That is,
more precisely, the density of eigenvalues of the Hessian of the
Hamiltonian, averaged over the equilibrium distribution.

The general frame where our computation acquires particular meaning,
and where a great number of recent analysis have been
performed, is the Instantaneous Normal Modes (INM) approach \cite{keyes}.
The main idea of this approach is that liquids are `solid-like' at
short times and that liquids' dynamics thus correspond  to vibrations
about  some equilibrium  positions with periodic jumps into new local 
minima \cite{inm}.

In this context, a  crucial quantity is the typical spectrum 
of the Hessian of the Hamiltonian, which
describes the structure of the energy landscape around
the typical configurations. This is  the INM spectrum 
and the related eigenvectors are the so called Instantaneous Normal Modes.
 
Recently there have been many attempts to relate quantitatively the
knowledge of the INM spectrum with measurable quantities, such as the
velocity-velocity correlation function, more complicated self-correlation 
functions, and even the diffusion properties \cite{keyes}. The main
idea is to generalize the standard harmonic analysis used for solids, taking
into account the finite hopping rate from one local minimum to the
other \cite{zwanzig,keyes}. 
For many systems simulations have shown that the predictions given by the
INM analysis are in very good agreement with the results obtained
directly from molecular dynamics. 
Simulations also show that the INM spectrum of a liquid system always
displays both positive and negative eigenvalues (i.e. real and
imaginary frequencies), not only in
the liquid, but also in the supercooled phase as the glass transition
is approached. While
the interpretation of the positive modes seems straightforward (they
represent harmonic vibrations in well defined wells), this is not the
case for the negative modes. The contributions to these modes come
from regions of negative curvature in the phase space sampled by the system
at equilibrium. Indeed the system does not remain forever in a local minimum
of the energy surface, but jumps into new regions over certain time
scales, in this way exploring an enormously complicated landscape.

Unfortunately, it is not possible to distinguish  in a simple way 
the modes related to real barriers (which can be  used to give 
an estimate of the hopping rates, see ref. \cite{keyes}) from the ones 
related to anharmonic  deformations of the local landscape. 
Recently it has been proposed
\cite{bembenek-loca}  that a crucial difference exists between
localized and extended negative modes: the localized modes involve a 
finite number of particles and can be associated to local 
barriers hopping; the extended
modes involve an extensive number of particles and
can be associated to structural rearrangements. Thus
these two kinds of modes should  be related to
different diffusional processes. Simulations seem to confirm this view
for some fragile models: in \cite{bembenek-loca} it is shown
that the number of extended negative modes goes to zero as a particular
temperature is approached and that this temperature is the same as the
temperature $T_c$ where the Mode Coupling Theory (MCT) would predict a
dynamical transition \cite{gotze}.
The temperature $T_c$ represents an important reference value for
fragile glasses: above $T_c$ MCT  successfully
predicts the observed dynamical behaviour of the system, and the
viscosity has a non-Arrhenius temperature dependence; below $T_c$ 
MCT breaks down and  the viscosity behaviour becomes Arrhenius-like.
It is commonly believed that $T_c$ coincides with the crossover
temperature postulated long ago by Goldstein \cite{goldstein}, below which
hopping processes become relevant. The result of \cite{bembenek-loca} 
shows that, for the particular system  studied there, 
$T_c$ can be found via the spectrum analysis, looking for
the temperature where the extended negative modes disappear.
It is not clear how general this phenomenon is, 
but it anyway indicates that the localization properties of the INM
represent an useful tool, and can be used to give at least an estimate
of the temperature $T_c$. 

For all these reasons, the analytic determination of the INM spectrum 
represents a vital task for any theoretical study of liquids. Indeed
there have been in the last years various attempts to perform such a
computation and some important steps in this direction have been done in 
\cite{wuloring} and \cite{strattI}.
In the following we will try to  outline what the standard
procedure to compute analytically the INM spectrum is, and we will
briefly mention what are the main assumptions and strategies adopted
in these previous works \cite{wuloring,strattI}. 
This will help us in introducing our new approach, 
commenting on what are the differences in perspective and procedure,
and what are the objectives we propose to address.

\section{General procedure}

Let us consider a system of $N$ interacting particles with Hamiltonian
\beq
H=\sum_{k>l}^N V(r_{kl}) \ ,
\eeq
where $V(r)$ is a two-body potential.  The Hessian matrix $\bf A$ is
defined by, $A_{kl}^{\mu\nu} = \partial_k^\mu \partial_l^\nu H$,
with $\mu, \nu=1,\dots,d$, being $d$ the dimension of the space.
The general form of $\bf A$ with respect to particle indices is
\beq
A_{kl} = -J_{kl}+\delta_{kl}\sum_i^N J_{ki}  \ , 
\label{j}
\eeq
where
$J_{kl} =  J({\vec r}_{kl})=V''(r_{kl}) {\hat r}_{kl} {\hat r}_{kl} + 
\frac{V'(r_{kl})}{r_{kl}}(1-{\hat r}_{kl} {\hat r}_{kl})$,
${\hat r}_{kl}$ being the versor along the inter-particle $k,l$ distance.  
The diagonal term of $\bf A$ is a consequence of the translational 
invariance of the system, which requires $\sum_k^N A_{kl}=0$. 

The standard procedure to compute the INM spectrum is to consider the
well known relation between the density of eigenvalues of the matrix
$\bf A$ and the correspondent resolvent operator ${\bf
G}(\lambda)=(\lambda {\bf 1} - {\bf A})^{-1}$:
\beq
D_A(\lambda)=\lim_{\epsilon \to 0} - \frac{1}{N \pi} 
{\rm Im}\  {\rm Tr}\  {\bf G}(\lambda - i  \epsilon) \ .
\label{start}
\eeq
This formula provides the spectrum of a single matrix ${\bf A}$, 
that is of the
Hessian matrix evaluated in a particular particles configuration. To
obtain the INM spectrum $D(\lambda)$, one has to average Eq. (\ref{start})
over the Boltzmann equilibrium distribution. In this way  
the computation of the INM
spectrum is reduced to the computation of the average diagonal element
of the resolvent matrix $\langle G_{ii}(\lambda-i  \epsilon)
\rangle$. The standard procedure is at this point to use an integral
representation for the resolvent:
$
\langle G_{ii}(\lambda -i\ \epsilon) \rangle 
= \frac{1}{Z}\int d\phi_1 \dots d\phi_N  \
\phi_i^2\ \exp[{-\frac{1}{2} \phi (\lambda - {\bf A} + i \epsilon)\phi}] \ .
$
In this way some  new variables, the internal fields $\phi_i$, enter
in the computation, beside the space coordinates $r_i$ included in
the explicit expression of the Hessian $\bf A$. 
A further step consists in adopting a replica trick to bring the
normalization factor $Z$ at the numerator:
$\frac{1}{Z}=\lim_{n\to 0} Z^{n-1} \ .$
In this way we  finally get:
\beq
\langle G_{ii}(\lambda -i\ \epsilon) \rangle
= \lim_{n\to 0} \int d {\vec\phi}_{i} \ (\phi^1_i)^2 \ \Omega({\vec\phi_i})
\label{inte}
\eeq
where now the ${\vec\phi_i}$ are vectors in a $n$-dimensional space.
The function $\Omega$ is given by
\beq
\Omega({\vec\phi}_i)=\int d{\vec\phi}_1 \dots d{\vec\phi}_N \ 
\langle e^{- \frac{1}{2}\vec\phi(\lambda-{\bf A}+i\ \epsilon)\vec\phi }
\rangle \ ,
\label{omega}
\eeq
where the integration is performed over all, but the $i$, internal fields. 

Of course the difficult task is to compute the
function $\Omega$. In \cite{wuloring,strattI}, more or less
explicitly, the authors assumed a Gaussian shape for $\Omega({\vec\phi})$, 
and then computed self-consistently its variance using a sort of
generalized liquid theory (we will come back to this point
later). Thus,   they assumed a quite rough approximation for the
general ${\vec\phi}$ dependence of $\Omega$, but took into account
accurately the many particles contribution terms included in the definition 
(\ref{omega}). Our approach has been precisely the opposite one. As
you will shortly see, we did not assume any  {\it a priori} form
for  $\Omega({\vec\phi})$, but we disregarded  many particles
contribution terms. As a result, we have an approach which is much
simpler to deal with, but still allows for non trivial spectral
properties. In \cite{wuloring,strattI}, excellent
results were obtained for the density of eigenvalues, but, on the
other hand, the computational procedure was too complicated to
reasonably look at the localization properties of the
eigenfunctions. Besides, a Gaussian shape for $\Omega$
is anyway  too simple to detect non trivial localization properties.
On the contrary, our simpler approach is more suitable for an eventual 
study of localization.

\section{Our approach}
We consider a liquid system at low density, that is
precisely in the physical context where we can expect many particle 
correlations not to be important. 
At low density the most part of the particles will be very far one
from the other and we can imagine that, picking out at random a couple
of particles $i$ and $j$, the interaction $V(r_{ij})$ between them
will be very  low, practically zero. 
This will be true also for the Hessian element
$A_{ij}$ between the two particles, since the Hessian is much more short
ranged than the potential itself. 
The whole matrix ${\bf A}$ will therefore have the most part of its
elements equal to zero and very few ones different from zero: it will be,
in other terms, a {\it diluted} matrix. 
In this context, the main idea underlying our approach is to
model the matrix ${\bf A}$ as a random diluted matrix 
and to use all the techniques developed for random matrix
ensembles to compute the INM spectrum.

As a first step in  this program, we must 
find out what is the probability  distribution of the matrix $\bf A$,
or, which is the same, of the matrix $\bf J$. 
This distribution is naturally induced  by the equilibrium 
probability over the positions of the particles via expression (\ref{j}). 
Consistently with our low density approximation, we can assume
that the probability distribution $P[{\bf J}]$ is factorized into 
the individual probabilities of the particles pairs. In this way the elements
of ${\bf J}$ (but not of $\bf A$) are independently distributed, i.e.
\beq
P[{\bf J}]\equiv\prod_{k>l}^N p(J_{kl}) \ .
\label{fact}
\eeq
It is clear that with (\ref{fact}) we are disregarding three-particles
correlations. 
Thus, (\ref{fact})
becomes a reasonable assumption when three-particles correlations are not 
important, for example at low densities.

Once assumed this factorized form for $P[{\bf J}]$, we can express 
the pair-probability $p(J_{kl})$ as,
\beqa
p(J_{kl}) & = & \int dr_1 \dots dr_N \; e^{-\beta H}
\delta( J_{kl}-J(r_{kl}) ) \non \\
& = &\frac{\rho}{N} \int dr_k  dr_l \; g^{(2)}(r_{kl})
\delta( J_{kl}-J(r_{kl}) )	
\label{uno}
\eeqa
where $g^{(2)}(r)$ is the two-particles correlation function and
$\rho$ is the average density. 
At this point there are various ways to exploit equation
(\ref{uno}). One can for example 
insert in (\ref{uno}) the numerical values obtained for $g^{(2)}(r)$
by numerical simulations, or by some classical liquid theory
approaches. Alternatively, and this is the simplest possibility, we
can adopt a low density expansion for the two points correlation 
function. In this way, $g^{(2)}(r)=\exp(-\beta V(r))$, at the first
order of the virial expansion.  This approximation is consistent with  
our previous assumption (\ref{fact}) where we disregarded
three-particles correlations, and,  as we shall see, it
enables to perform completely analytically the calculations.

In the general frame we have described, we finally have the Hessian
matrix $\bf A$, and its distribution, as given by (\ref{uno}). This is
enough to start a random matrix computation. However, the algebra is
still very complicated: the matrices involved ($\bf A$ and $\bf J$) are
actually tensors with particle and space coordinates indices. Before
dealing with this general case, we decided to test our method under the
simplest possible conditions, that is when the particles live in one dimension.
This case is conceptually analogous to the three dimensional one, with
the advantage of a simpler algebra.

\section{The one dimensional case}

In one dimension the explicit expression of the matrix $\bf J$ is much
simpler, since only the longitudinal part of (\ref{j}) survives, and
we have $J(r)=V''(r)$. To obtain the distribution $p(J_{kl})$ we have
to solve equation (\ref{uno}). In the light of the greatest
simplicity we consider a  soft-spheres potential $V(r)=1/r^m$. 
Thus, from eq.(\ref{uno}) we have, 
\beq
p(J) \sim \frac{1}{N}\; \frac{e^{-\hat\beta J^{b}}}{J^{1+c}}\equiv
	\frac{1}{N} \; q(J) \ ,
	\label{qj}
\eeq
with $\hat\beta=\beta \; [m(m+1)]^\frac{1}{m+2}$, $b=m/(m+2)$
and $c=1/(m+2)$. (From now on we indicate with $J$ an individual
element of the matrix $\bf J$). 
For realistic values of $m$ (typically $m=12$) the
parameter $b$ is very close to one. Therefore, we will directly set
$b=1$ in $p(J)$ in order to simplify our calculation. 
We will show in the discussion of our results that the actual 
spectrum is very weakly dependent on this approximation.

As it stands the distribution $p(J)$ is not normalizable, but we can
regularize it in the following way. Let us put an IR cut-off $\bar r$, 
by setting $V(r)=0$ for $r>\bar r$, and let $\eta=V(\bar r)$. We obtain in
this way a regularized form of the pair probability:
\beq
p_\eta(J) = \delta(J) + \frac{1}{N} 
\left( q(J)\; \theta(J-\eta) - 
\delta(J) \int_\eta^\infty dJ' \; q(J') \right) \ , 
	\label{pj}
\eeq
where $q(J)$ is defined in equation (\ref{qj}).
A few comments about equation (\ref{pj}):
\begin{itemize}
\item
the distribution  $p_\eta(J)$ is  {\it diluted} as we expected, 
since the probability of finding an element of the matrix $\bf J$
equal to zero is of order one, while the probability of finding 
one element larger than $\eta$ is of order $1/N$.
\item
$p_\eta(J)$ explicitly depends on the value $\eta$ of the
cut-off. However, if we consider a generic function $f(J)$ and compute
its average over $p_\eta(J)$, we get:
$
\langle f(J) \rangle \equiv \int_\eta^\infty dJ \; q(J) \; [f(J)-f(0)]  
$.
Thus, if the function $f$ is differentiable in $0$, its
average value has a well defined limit when $\eta \to 0$. This means
that in our computation, after averaging over the Hessian
distribution, we can safely take the limit $\eta \to 0$, recovering
the original problem without cut-off.
\end{itemize}

At this point we finally have a well defined random matrix problem: we
have an ensemble of matrices $\bf A$, and their distribution, as given
by equation  (\ref{pj}). We can therefore try to apply random matrix
techniques to compute the density of eigenvalues $D(\lambda)$.

Before going on with the computation,  we would like to add some
more general remarks on the one dimensional case. 

First of all, we
note that the low density approximation we made when disregarding 
three-particles correlations in (\ref{fact}) is, at fixed density and
temperature, the  less appropriate the lower  the dimension. For
$d=1$ the geometrical constraints on the particles are much stronger
and indeed the true Hessian matrix has a band structure. For this
reason our analysis has to be regarded more as a training example for the
three dimensional case rather than  a predictive computation for a real one
dimensional system (our one dimensional treatment is very similar to
what, for an electronic band structure problem, would be an $s$-band  
computation \cite{strattII}). 

Secondly, in the specific example we are
dealing with we have chosen a soft-sphere potential $V(r)$. It is
easy to see that in this case the Hessian $\bf A$ is a positive
defined matrix which therefore has a positive defined density of 
eigenvalues. This is  not the case for a three
dimensional system, where the role of the negative modes is an important 
issue.  However, the qualitative
shape of the spectrum is actually very similar to what found, for example, in
simulations on three dimensional systems \cite{srinuovo}. 
Moreover, the eigenfunctions exhibit  non
trivial localization properties in the tails, thus providing a
good context where testing  a localization analysis procedure.
Finally, to avoid confusion in the future, we note once again that the matrix
$J$ is {\it not} a nearest-neighbors matrix, but a diluted one. This
is why, even in one dimension, we find both localized and extended
states (and not only localized ones \cite{review}).

Let us now proceed with the computation.
The procedure is very 
similar to the one outlined in the previous section for the standard
computation of the INM spectrum: one has to relate $D(\lambda)$ to the
resolvent operator ${\bf G}(\lambda)$, and the diagonal element of the
resolvent to the one-particle function $\Omega$. The main difference is
that now all the averages involved in the  definition (\ref{omega}) 
of  $\Omega$ are not averages over the Boltzmann distribution, but  over
the distribution  $p_\eta(J)$.

The crucial point is, of course, the computation of
$\Omega(\vec\phi_i)$. This can be done in different ways, generalizing
some random matrix computations \cite{diluite} to the particular 
case of the distribution $p_\eta(J)$ \cite{noispettro}. 
In our case it is convenient to write $\Omega$ as (the particle index
$i$ is, from now on, understood)
$
\Omega(\vec\phi)= e^{-\frac{1}{2}\lambda \phi^2 + g(\vec\phi)}
$
and look for a self consistent equation for the exponent $g(\vec\phi)$.
We note that $g(\vec\phi)$ measures how much the function $\Omega$ is
Gaussian: a non-quadratic shape of $g$ implies a non-Gaussian $\Omega$.

We will not enter into the details of how obtaining the self
consistent equation (the interested reader is referred to
\cite{noispettro}), but simply give the result: 
\beq
g(\vec\phi)=\int d\vec\phi' \; e^{-\frac{1}{2}\lambda
\vec\phi'^2+g(\vec\phi')}\int dJ \; q(J) \left [ e^{-\frac{1}{2} J
(\vec\phi-\vec\phi')^2 }-1 \right ]
\label{nonav}
\eeq
This equation has still to be averaged over the distribution
$q(J)$. This can be done exactly with a few algebraic tricks, and
finally we get 
\beq
{\hat g}(x)= g(e^{i\pi/4}x)= K_1(x)
 - x\int_0^\infty dy \; K_2(x,y) \; \exp(i\frac{\lambda}{2}y^2 + {\hat g}(y))
	\label{lei}
\eeq
where $x=|\vec\phi|$, and $K_1(x)$ and  $K_2(x,y)$ are 
expressed in an analytic form \cite{noispettro}.

First of all, it is possible to check analytically that asymptotically
$g(x)\sim x^{2 c}$, thus proving that $\Omega$ is definitely {\it not}
a Gaussian function. Besides,  we have been able to numerically 
solve the equation for $g(x)$ without any further approximation. 
Indeed, eq.(\ref{lei}) has the form of a fixed-point
equation and can be solved  numerically  by iteration, discretizing
the function $\hat g$ and the kernel $K$ on a lattice. 

Once obtained $g$ for a given value of $\lambda$, it is possible 
to compute  the spectrum, using backward the relations between
$\Omega$ and the resolvent, and finally between the resolvent and $D(\lambda)$
(see previous sections).
The results are shown in Fig.1a, where we have plotted the INM spectrum
$D$ as a  function of $\lambda$, for $m=12$.
The spectrum has positive support and it depends on the
scaled inverse temperature $\hat\beta$ in the expected way: for low
temperatures (high $\hat\beta$) the collisions among particles 
are weaker, so that the spectrum is peaked on lower values of the
eigenvalues. On the other hand, the tail for large $\lambda$ is 
larger at higher temperature. The behaviour in the right tail is
of the form $D(\lambda) \sim e^{-\alpha \lambda}$, where $\alpha$ is
an increasing function of $\beta$.

As previously stated, this one dimensional case represents for us,
first of all, a way to test our analytic procedure.  
A crucial task is therefore to check whether the result we have 
found is correct. To this aim we have done extensive numerical 
simulations. Once drawn a matrix $\bf J$ with probability (\ref{pj}), 
we build $\bf A$ and diagonalize it numerically. 
Since the spectrum has huge tails for large eigenvalues, it is 
convenient, in order to compare simulations with analytic 
results, to consider the probability distribution $\pi$ of 
$\mu\equiv\ln \lambda$, that is $\pi(\mu)=D(e^\mu) e^\mu$. 
In Fig.1b we plot $\pi(\mu)$ as obtained from the analytic form of 
$D(\lambda)$, together with the one obtained from numerical simulations.
The two curves are in excellent agreement confirming the validity of 
our result. Besides, we show in the inset of Fig.1b the numerical 
spectrum obtained with the original value of 
$b=m/(m+2)$. 
The result justifies the sensibility of the approximation $b\sim 1$.

In the introduction we mentioned that an important issue in the context
of the INM calculations is the analysis of the localization properties
of the negative modes. As we have previously said, 
even if in this one dimensional case the spectrum is positive defined,  the
tails exhibit non trivial localization properties. 
From a numerical point of view an important quantity to investigate
localization properties is the average inverse 
participation ratio $Y(\lambda_\alpha)=\sum_{i=1}^N 
\left(w_\alpha^i\right)^2$, 
where $\alpha=1\dots N$ is the eigenvalue index and 
$w_\alpha^i=[\langle \alpha | i \rangle]^2 $ is the weight of
site $i$ in the eigenfunction $| \lambda_\alpha \rangle$. 
In Fig.2 we plot $Y$ as a function of $\lambda$, as obtained via numerical
diagonalization. 
It is clear from the figure that there are two localizations edges, 
separating a central region of extended eigenvalues, from the 
tails where localized states are present. 
Note that for $\lambda\to 0$ the inverse participation ratio goes to
one and this corresponds to a single particle which happens to 
be nearly decoupled from the rest of the system. 
On the other hand, the localized states of the right tail correspond
to pairs of very strongly interacting particles and this naturally
leads to a inverse participation ratio equal to $1/2$.

Our aim for the future is  to find analytically the two localization edges
revealed by the numerical analysis. Our idea is to generalize to
our case the methods used for the Bethe Lattice 
\cite{loca} and we will address this problem in 
a future work  \cite{noiprepa}.

\section{Comparison with other approaches}
In order to better understand what are the differences between our
approach and the ones of \cite{wuloring,strattI}, it is convenient to
go back once again to the general procedure outlined in section II,
and,  in particular, to expressions (\ref{inte}) and (\ref{omega}). A
crucial observation made in \cite{strattI} is that the function
$\Omega(\vec\phi)$ is proportional to the one-particle correlation function
of a generalized liquid system, whose coordinates are
$X_i=(r_i,\vec\phi_i)$ and whose Hamiltonian is given by:
\beq
-\beta H_{GEN} = \sum_i U(X_i) + \sum_{i<j} W(X_i,X_j) \ ,
\label{hgen}
\eeq
where $U(X_i)=-\frac{1}{2} \lambda \ (\vec\phi_i \cdot \vec\phi_i)$ and 
$W(X_i,X_j) =-\beta V(r_{ij})+ A_{ij}\ (\vec\phi_i \cdot \vec\phi_j)$.
Indeed it is simple to check that
$
\Omega(\vec\phi)=\langle \frac{1}{N}\sum_i \delta(\vec\phi-\vec\phi_i)
\rangle_{GEN} = s(X)/\rho$,
where $s(X)$ is the  one-particle correlation function of the
generalized liquid system, and $\rho$ is the average density of the
original system.
Given this analogy, it is possible to compute $\Omega$ using
the standard liquid theory \cite{hansen} applied to this generalized 
liquid system.
This is precisely what the authors of \cite{strattI} did: they assumed
a Gaussian shape for $\Omega$, and then computed the variance using a
renormalized Mean Spherical Approximation \cite{hansen,morita,class}. 

An alternative route is the following. We can use a Hypernetted Chain
(HNC) approximation \cite{hansen,morita,class} to obtain the one-particle
correlation function of the generalized liquid. The HNC equations can
be deduced from a variational principle where the free energy of the
system is written as a functional of the one-particle and two-particles
correlation functions \cite{morita}. The variational free energy has
the following form:
\beqa
 \beta F = \frac{1}{2} \int dX dX' s(X) s(X') \ g^{(2)}(X,X')
\log(g^{(2)}(X,X')) +\frac{1}{2} \int dX dX' \non\\ 
s(X)  s(X') \left [1- g^{(2)}(X,X')+\beta W(X,X') \ g^{(2)}(X,X') \right ] -
\int dX s(X)U(X)   \non\\
 + \int dX s(X) [\log(s(X))-1] -\frac{1}{N}{\rm Tr}\left [\log(1+s h)-s h +
\frac{1}{2} (h s h s ) \right ]  \ ,
\label{elibera}
\eeqa
where $s(X)$ and $g^{(2)}(X,X')$ are, respectively, the one-particle and the
two-particles correlation functions, $h(X,X')=g^{(2)}(X,X')-1$ and in the
trace term all the products are convolutions. 
Self-consistent equations for $s(X)$ and $g^{(2)}(X,X')$ are obtained
variationally from (\ref{elibera}).
These equations are in general not easy to solve, but one can 
start facing them perturbatively. For example, we can look at a
 low density situation where, in a first approximation, the trace term will
not contribute. In this case, the variational equation for the two-particles
correlation function gives immediately
$
g^{(2)}(X,X')=e^{-\beta W(X,X')} \ ,
$
which is nothing else than the first order of the virial expansion.
The equation for the one-particle correlation function is, on the other
hand, less trivial. In terms of  $\Omega(\vec\phi)=s(X)/\rho$ it gives
\beq
\log(\Omega(\vec\phi))+ \frac{1}{2}\ \lambda \ \vec\phi^2 =
\int d\vec\phi' \; \Omega(\vec\phi') \rho \left [ \int dr
e^{-\beta V(r)} e^{\frac{1}{2} J(r)(\vec\phi-\vec\phi')^2}-N \right ]  \ .
\label{nostra}
\eeq
The interesting fact about equation (\ref{nostra}) is that it is
exactly the same self-consistent equation that we get in our low
density approach presented in the previous section.
To see this, we note that
$
\rho \int dr e^{-\beta V(r)} 
e^{\frac{1}{2} J(r)(\vec\phi-\vec\phi')^2} = \int dJ q(J) 
\left [ e^{\frac{1}{2} J (\vec\phi-\vec\phi')^2 }-1\right ]+ N  
$.
If we use now the previously adopted notation  
$\Omega(\vec\phi)=\exp(-1/2\ \lambda \vec\phi^2+g(\vec\phi))$ and  we insert 
it in (\ref{nostra}) we finally get precisely equation (\ref{nonav})
for $g(\vec\phi)$.

This result shows that our simple low density approach can be seen as
the first order approximation of a more complicated
generalized HNC approach. We understand now in a deeper way what is the
origin of the approximations we made and also what is the route we
have to follow to improve our calculation. Indeed, we see that, if we
want to go beyond the low density approximation, we have  to
consider further terms in the trace appearing in (\ref{elibera}) and
compute consequently the self-consistent equations  \cite{noiprepa}. 

\section{Conclusion}
We have presented a new approach for the analytic computation of the
INM spectrum. At present we have successfully applied
this approach to the case where particles live in one dimension, and at 
low density. This simple case represents the ideal context where understanding
the potentiality of our method and outlines the procedure one has to
follow in the more general cases. 

We expect this approach to  give good results in the realistic 
three-dimensional situations, where the geometrical constraints are much 
less important than in one dimension. A strong  indication in this direction
is provided  in Fig.2b, where we plot the spectrum obtained from a Montecarlo
simulation for a soft-sphere mixture. The thick curve is obtained considering,
for each sampled configuration, the real Hessian matrix $\bf A$ of the system.
The  points, on the other hand, are obtained considering a 'scrambled' Hessian 
matrix where, for each sampled configuration, a  new $\bf J$ matrix is 
built by mixing at random the elements $J_{ij}$ of the real $\bf J$.
This scrambling procedure destroys the three-particles correlations and  
is therefore equivalent to our approximation (\ref{fact}).  The figure shows
clearly that the scrambled approximation works very well for 
$\Gamma=\rho\beta^{1/4}=0.2$. A more systematic analysis \cite{noiprepa} 
with Montecarlo simulations indicates that good results are obtained also 
at much higher values of $\Gamma$ suggesting that our analytic approach, 
even at the simplest first step where three-particles correlations are 
discarded, should describe quantitatively well the real spectral properties
of a three-dimensional system on a wide range of temperatures and densities.

\section{References}


\eject

\begin{figure} 
\hbox to\hsize{
\epsfxsize=0.5\hsize\hfil\epsfbox{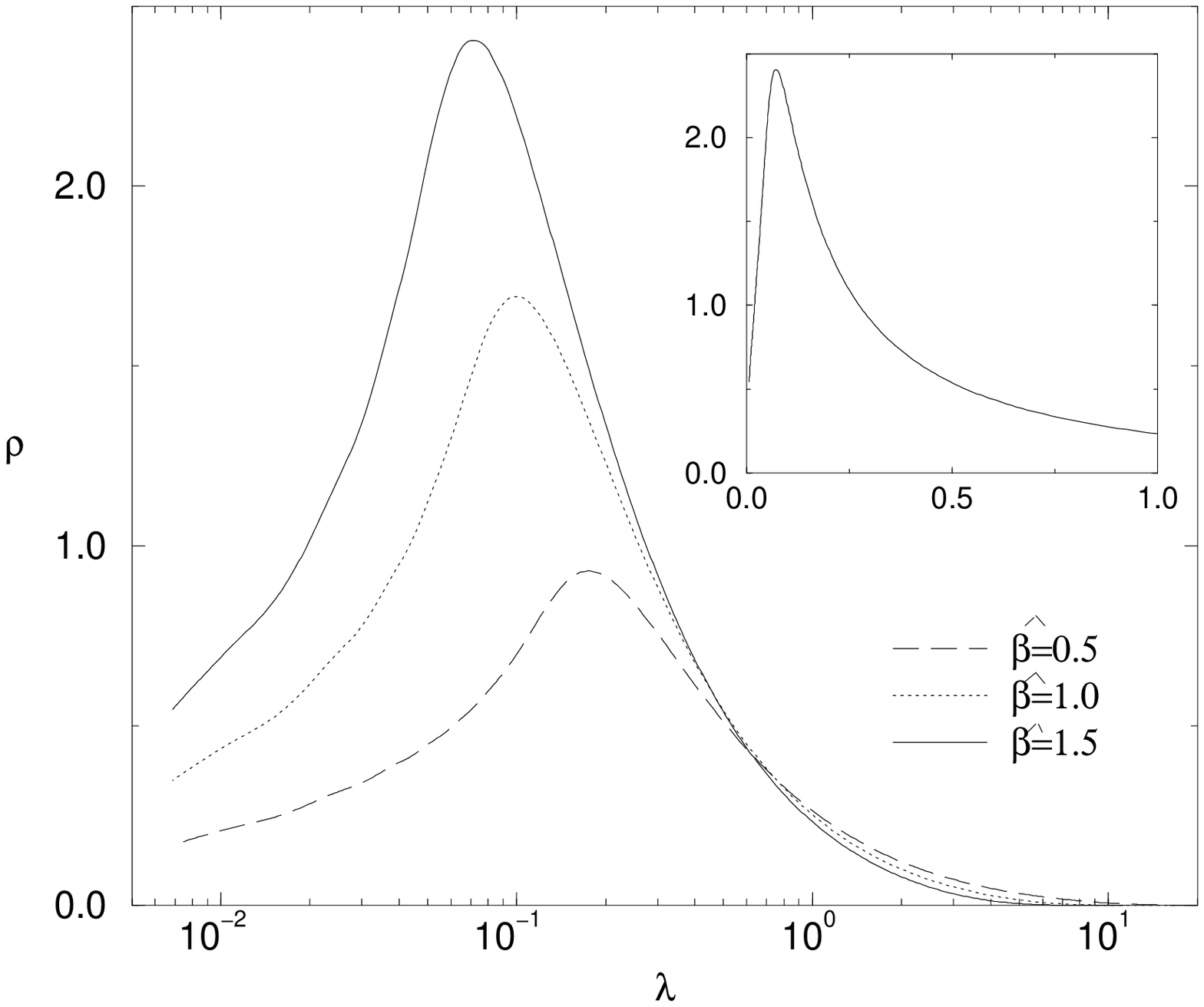}
\epsfxsize=0.5\hsize\hfil\epsfbox{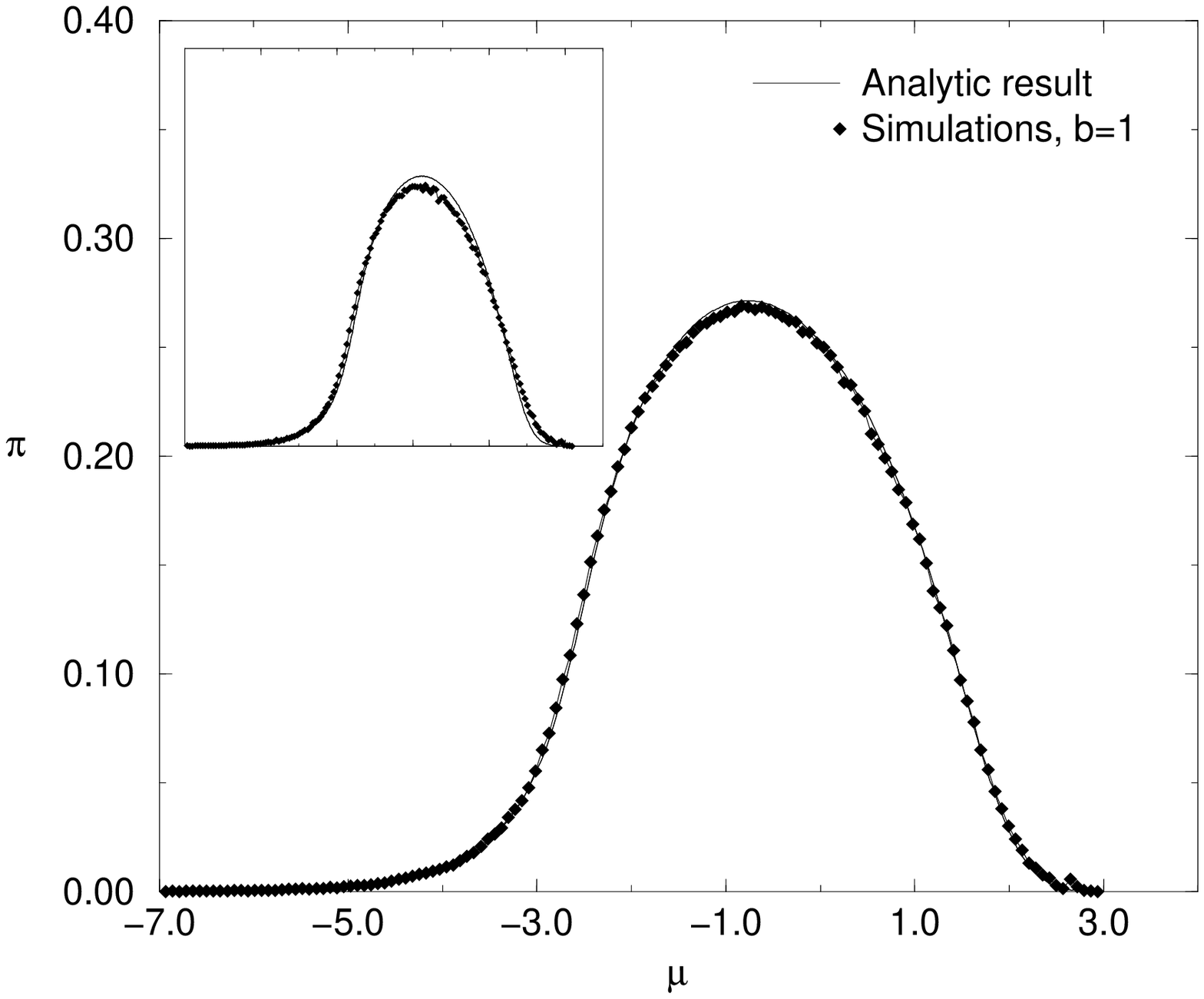}
\hfil}
\caption{
(a) The INM spectrum $D$ as a function of $\lambda$ for
different values of the scaled 
temperature $\hat\beta$, $m=12$ and $\epsilon=0$. 
The plot is in log-linear scale. In the inset it is shown 
$D(\lambda)$ for $\hat\beta=1.5$ in linear-linear scale. 
The spectrum vanishes at $\lambda=0$.
(b) Numerical simulations vs. analytic solution. 
We plot here the probability distribution $\pi(\mu)$, with $\mu=\ln \lambda$. 
Both the curves correspond to a probability distribution with $b=1$. 
$N=600$ and $\eta=10^{-4}$. $\hat\beta=1$ and $m=12$.
In the inset, on the same scale, we compare the analytic
result for $b=1$ with the simulations performed with $b=m/(m+2)$.
}
\label{fig1} 
\end{figure} 

\begin{figure} 
\hbox to\hsize{
\epsfxsize=0.5\hsize\hfil\epsfbox{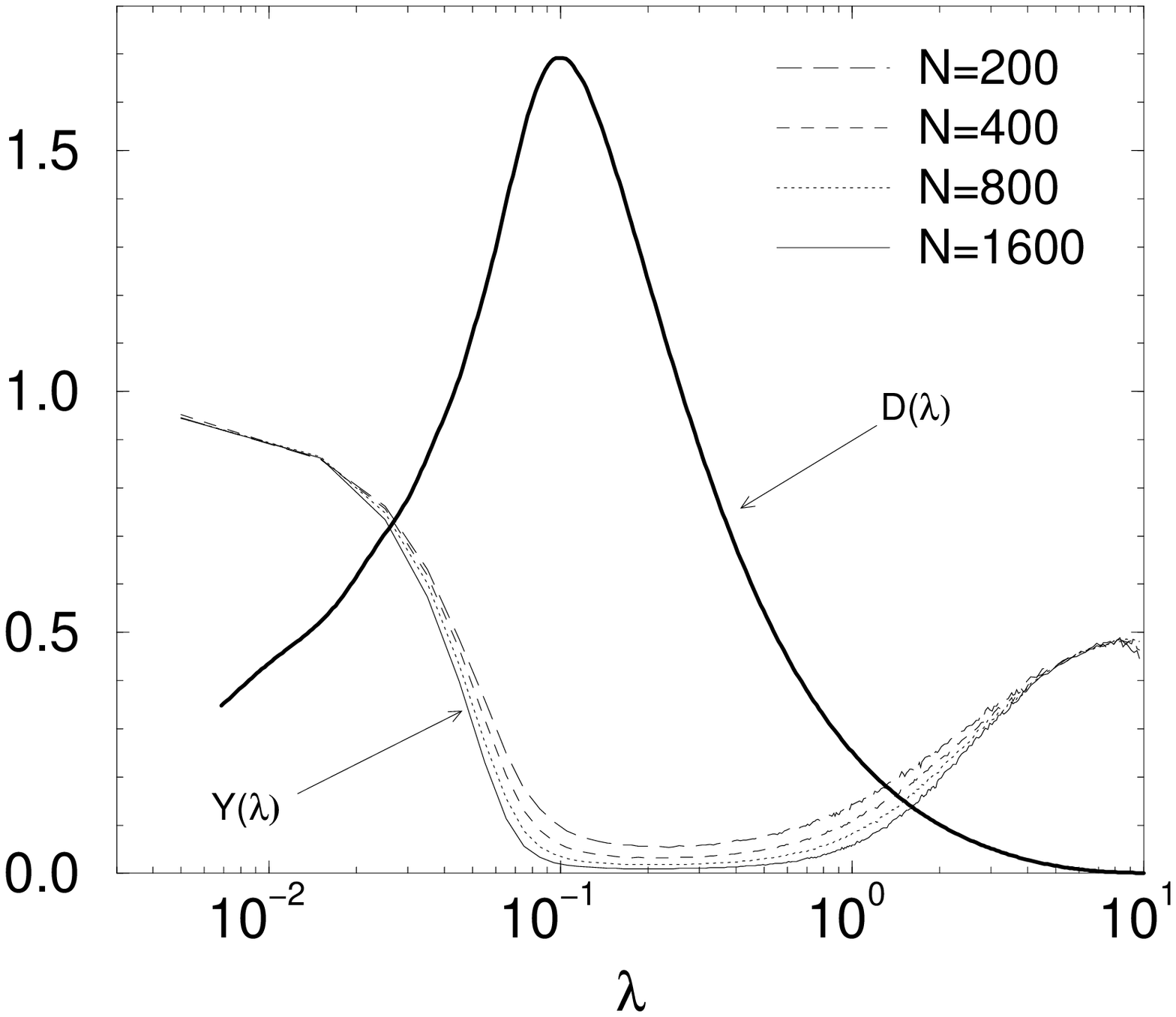}
\epsfxsize=0.5\hsize\hfil\epsfbox{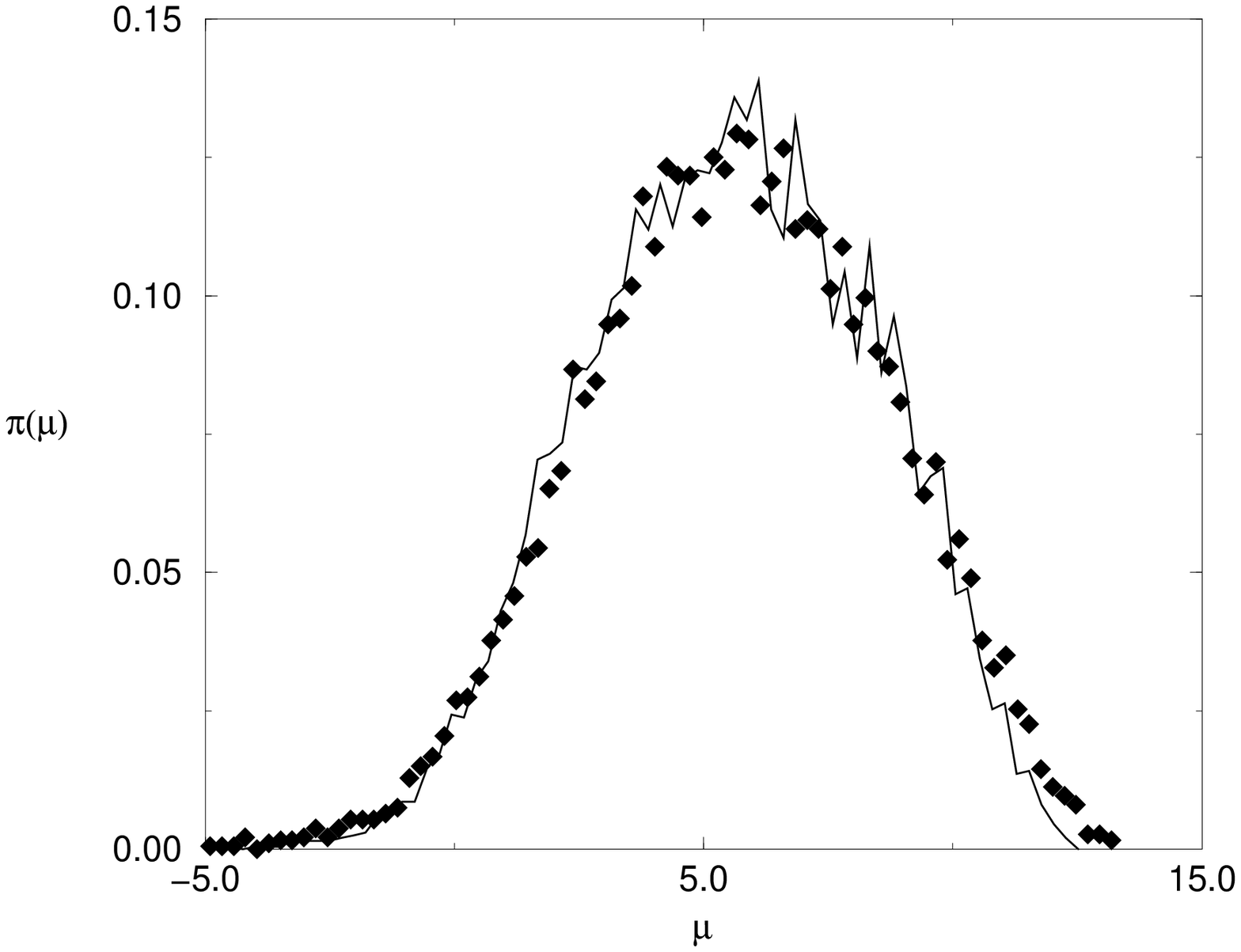}
\hfil}
\caption{
(a) Inverse participation ratio $Y$ as a function of the
eigenvalue $\lambda$ at different values of $N$. $\eta=10^{-4}$, 
$\hat\beta=1$ and $m=12$. The thick curve is the spectrum $D(\lambda)$
at the same values of the parameters.
(b) Montecarlo simulation for a three-dimensional soft-sphere mixture, at 
$\Gamma=0.2$. The thick curve represents the real INM spectrum. The points 
correspond to the spectrum obtained using a scrambled Hessian matrix.  
}
\label{fig2} 
\end{figure}

\end{document}